\documentclass[12pt,showpacs,preprintnumbers,amsmath,amssymb,nofootinbib]{revtex4}
\usepackage{color}
\usepackage{latexsym,epsf}
\usepackage{graphicx}
\usepackage{bm}
\usepackage{epsfig}
\begin{document}
\preprint{{\vbox{\hbox{NCU-HEP-k024}
\hbox{Mar 2006}\hbox{rev. Nov 2006}\hbox{ed. Feb 2007}
}}}
\relax
\def\ga{\mathrel{\raise.3ex\hbox{$>$\kern-.75em\lower1ex\hbox{$\sim$}}}}
\def\la{\mathrel{\raise.3ex\hbox{$<$\kern-.75em\lower1ex\hbox{$\sim$}}}}
\renewcommand{\arraystretch}{1.2}
\newcommand{\be}{\begin{equation}}
\newcommand{\ee}{\end{equation}}
\newcommand{\bea}{\begin{eqnarray}}
\newcommand{\eea}{\end{eqnarray}}
\newcommand{\ba}{\begin{array}}
\newcommand{\ea}{\end{array}}
\newcommand{\bc}{\begin{center}}
\newcommand{\ec}{\end{center}}
\newcommand{\ssc}{\scriptscriptstyle}

\def\bsg{$b \to s \, \gamma\,$}

\vspace*{0.5in}
\title{Flavor Changing Higgs Decays in Supersymmetry with Minimal Flavor Violation\\[.5in]}
\author{\bf Abdesslam Arhrib$^{1}$, Dilip K. Ghosh$^{2}$, Otto C. W. Kong$^{3}$, and Rishikesh D. Vaidya$^{4}$}

\affiliation{\vspace*{0.5in}$^{1}$D\'epartement de Math\'ematiques, Facult\'e des Sciences et Techniques,
B.P 416 Tanger, Morocco\footnote{Permanent Address.}\\
and Physics Division, National Center for Theoretical Sciences, P.O. Box 2-131 Hsinchu, Taiwan.\\
$^{2}$Department of Physics $\&$ Astrophysics, University of Delhi, Delhi  110007, India.\\
$^{3}$Department of Physics National Central University, Chung-li,  32054, Taiwan.\\
$^{4}$Department of Theoretical Physics, Tata Institute of Fundamental Research, Mumbai  400 005, India.\\
}

\begin{abstract}
We study the flavor changing neutral current decays of the MSSM Higgs
bosons into strange and bottom quarks. {We focus on a scenario
of minimum flavor violation here, namely} only that induced by the CKM matrix.
Taking into account constraint from $b\to s\,\gamma$, $\delta\rho$  as well as
experimental constraints on the MSSM spectrum, we show that the branching
ratio of  {$(\Phi\to b\bar{s})$ and $(\Phi \to \bar{b}s)$ combined,
for $\Phi$ being either one of the CP even Higgs states,} can {reach}
the order $10^{-4}$-$10^{-3}$ for large $\tan\!\beta$, large $\mu$, and large
$A_t$. {The result illustrates the significance of minimal flavor
violation scenario which can induce competitive branching fraction for
flavor changing Higgs decays. This can be compared with the previous
studies where similar branching fraction has been reported, but with
additional sources of flavor violations in squark mass matrices.}
{We also discuss some basic features of the flavor violating decays in the generic case.}
\end{abstract}
\maketitle

\newpage
Supersymmetry (SUSY) is without doubt the most popular extension of
the Standard Model (SM). Its possible discovery has been a major target
of the Large Hadron Collider (LHC) program. Direct observation of some
of the superparticles as well as indirect hints through SUSY enhancement of
various rare processes in the SM are of great interest. Among the latter,
processes characterized by flavor changing neutral currents (FCNCs) are
particularly important. Especially within the framework of the minimal
supersymmetric standard model (MSSM), the $b \to s \, \gamma$
transition is an important case example that has received the most attention
\cite{mssm,DnH,KV}. Recently, there have been some studies on the related
flavor changing Higgs decays (FCHDs)
$\Phi \to bs$ ({\it i.e.} $\Phi \to b\bar{s}$ and $\Phi \to \bar{b}{s}$ combined;
$\Phi$ in the case denotes $h^{\ssc 0}$, $H^{\ssc 0}$, or $A^{\ssc 0}$)
\cite{1CT,2CH,3B,4H,5H}.
The basic results indicate that good branching fractions for various Higgs
states decaying into a $b$ and $s$ pair are admissible in some region of the
parameter space surviving the experimental $B \to X_s + \gamma$ constraint.
We consider careful, in depth, studies of the topic of great interest.
In this brief letter, we report one step in the direction, pointing out that even for the
case of a truly minimal flavor violation framework,  substantial
branching ratio is still admissible.  By the minimal framework, we mean
switching off all flavor off-diagonal squark mixings and use the CKM mixings
as the sole source of flavor violation. The result is apparently outside the
expectation of the previous authors. In fact, we find that the branching ratio
can reach the $10^{-4}$-$10^{-3}$ level for large $\tan\!\beta$, large $\mu$,
and large $A_t$, after the various relevant experimental constraints are put in.
We will also briefly review what has been done {in the generic case}
and discuss some related issues and suggestions for further studies.

Within the SM, fermions of the same charge have identical gauge interactions
and there is only one Higgs doublet giving mass to all the fermions through
various different Yukawa couplings. FCNC does not exist at tree level.
Presence of more than one Higgs doublet can in general lead to FCNCs
mediated by the physical Higgs states.
Large FCNCs can be avoided in the so called ``type-II'' two Higgs doublet
model, where one Higgs doublet $H_u$ gives mass to only the up-type quarks
and the other Higgs doublet $H_d$ gives mass to only down-type quarks and
charged leptons. Such a structure is protected against quantum corrections
by a discrete symmetry under which the two Higgs doublets transform differently.
The MSSM as a two Higgs doublet model has the type-II structure enforced,
however, only through the holomorphy of the superpotential. This obviously
cannot be effective once the soft SUSY breaking terms are taken into consideration.
Flavor off-diagonal mass mixings among the squarks is a potentially very
significant source of FCNC among quarks, starting at one loop level. Admitting
R-parity violation of course introduces further sources of flavor violation {\cite{rpv}}.
We are, however, limiting ourselves to the case with a conserved R parity here.
On the other hand, the nontrivial CKM matrix  does give rise to FCNC processes
through violation of the famous GIM cancellation among the charged current loops.
This minimal, but unavoidable, flavor violation does induce extra SUSY
contributions to FCNCs among quarks. In particular, it does contribute
to  $b \to s \, \gamma$ and $\Phi \to bs$.

Among the various FCNC processes,  $b \to s \, \gamma$ is arguably the most
extensively studied. In spite of the inherent
complications from strong interactions, the inclusive $B \to X_s + \gamma$
result admits a relatively clean calculation that can be freed from
scale and scheme uncertainties within the framework of RG-improved
perturbation theory and heavy mass expansion based on the assumption
of quark-hadron duality. The SM calculation has been completed to next
to leading log precision\cite{sm} and the result stands in good agreement
with the experimental numbers.
Whereas, FCNCs in B decays have been extensively analyzed, not so much
attention has been given to flavor changing interactions of Higgs boson.
In fact, Higgs decays, being without complications that afflict B-decays, could
provide fertile ground for explorations of FCNCs.  This is especially the
case for $\Phi \to bs$ in the large  $\tan\!\beta$ domain. Such an investigation
could also be interesting from the perspective of Higgs discovery. For
example, it has been pointed out in Ref.\cite{bejar-LHC}
that the total number of FCNC heavy flavor events from supersymmetric
Higgs boson interactions at the LHC can be as large as $\sim 10^6$. In
the critical range for the lightest CP even Higgs mass $m_{h^0} \sim
90-130\,\mbox{GeV}$, {the dominant decay mode $h^0\to b\,\bar b$ in large
$\tan\!\beta$ region} suffers huge QCD background. It even loses the race when
compared to the cleaner signature of $h^0 \to \gamma \gamma$, which has a much
smaller width. The FCNC decay mode, being free from pure QCD background
could {possibly} play a complementary role in identifying an MSSM CP even
Higgs{\cite{bob}}. Another interesting property of the FCHDs is their
non-decoupling nature in the limit of heavy SUSY spectrum limit\cite{nonD}.
In a pessimistic scenario where the superparticles  are heavy enough to evade
direct discovery at the next generation collider experiments, the non-decoupling
behavior would then play a very important role in revealing SUSY.

A  full diagrammatic calculation on the SM branching fraction for  $h \to bs$
has been performed in Ref.\cite{abdes-2HDM}. The results are $10^{-7} \,(10^{-9})$
for $m_h = 100\,(200)\,$GeV, in basic agreement with the earlier estimates
based on dimensional analysis and power counting\cite{3B,bejar-2HDM}.
Ref.\cite{abdes-2HDM} has also given a systematic analysis on the topic for
the case of the two Higgs doublet model (without SUSY). For the type-I case, it has
been found that the branching ratio falls in the range $10^{-5}-10^{-3}$ for small
$\tan\!\beta \approx 0.1-0.5$; for the type-II model with the \bsg constraint imposed,
$10^{-5}-10^{-6}$ can be obtained.  The CKM-induced, or SM, part of the results
is dramatically larger than the up-sector counterpart, {\em i.e.} Br$(h\to {{t}}c)$,
of $10^{-13}$\cite{sm1}. The down-sector case has two major advantages over
that of the up-sector. Firstly, the large mass of the top inside the loop gives a bigger
GIM violation. In addition, the $h \to bs$ channels are open even for
$m_h\la 2 M_{\!\ssc W}$ while the $h\to {{t}}c$ threshold of $m_t+m_c$ is about
or above the Higgs to $W^+W^-$ and  $Z^0Z^0$ channels. While the channels
with {virtual  W- or Z-boson(s)} do contribute, for Higgs mass quite a bit
below $2M_{\!\ssc W}$, the FCHDs may enjoy a much better branching
ratio. We take the opportunity to emphasize here that more favorable branching
ratio for the FCHDs in general, and $\Phi \to bs$ in particular, are likely to be
obtained where the widths of the usually dominating flavor conserving channels
are suppressed. Below the $W^+W^-$ threshold, $b \bar{b}$ is the dominating channel.

The MSSM is our focus model in this brief letter.  For the
present purpose, a good model background reference is provided by Ref.\cite{D}.
Our minimal flavor violation scenario means to focus on the
absolutely unavoidable CKM-induced results. They are from the charged current
loops involving  the W-boson, the charged Higgs, or the charginos. Only the loop diagrams
with the charginos are the truly SUSY contributions beyond what is available in the two
Higgs doublet model. They include diagrams with a (Higgs to) chargino vertex or with a
sfermion, most importantly the stop, vertex.

In the more general case, the soft SUSY breaking sector has many parameters
admitting flavor violation. Taking MSSM as a low energy effective theory
without assuming any particular model of SUSY breaking/mediation, the
parameters would be simply generic phenomenological parameters. On the other
hand, most of the available models of SUSY breaking/mediation assume a
flavor-blind mechanism, reducing the source of flavor violation to a minimum.
The latter is typically induced through {the flavor non-universal Yukawa
couplings and the RG-running due to the disparity between the involved scales.} This has
a preferential effectiveness on the terms involving the third generation and those
with $L$-handed sfermions, say the so-called $LL$-mixings over $RR$-mixings.
The $LR$-mixings introduced through the trilinear $A$-terms may also have
important flavor off-diagonal results. In fact, the $A$-terms themselves are
direct Higgs-sfermion couplings. {Such flavor off-diagonal couplings
of course induce FCHDs, especially through a gluino diagram with the strong QCD coupling
vertices.}

Flavor violation in the soft SUSY breaking sector adds to the CKM-induced
effect of the charged current loops. In addition, there are new contributions from the
gluino and neutralino loops. {Again,} the former is particularly
important due to the strong QCD coupling. Hence, with significant soft sector flavor
violation, the gluino-sfermion loops will dominate the FCHDs, while for small
or vanishing soft sector flavor violation, the chargino-sfermion loops  {should
be more important}. We would also like to point out that the $LL$-mixings
among the squarks of the up- and down-sector originate from the same squark doublet
mass terms. They are not identical, but related through a unitary rotation by the CKM
matrix. Hence, only with a doublet squark $\widetilde{Q}$ soft mass matrix
proportional to the identity matrix could the $LL$-mixings be completely removed.
We stick to this true minimal flavor violation case for our numerical study here.

A real Higgs particle still awaits discovery while a $b$-$s$ type flavor
violation beyond what is predicted within the SM is already stringently
constrained, especially in the very good agreement of the SM
$B \to X_s + \gamma$ number with the experimental results ---
the world average is given by \cite{pdg4}
\be
\label{br-exp}
Br\left[B \rightarrow X_s + \gamma \;(E_{\gamma} > 1.6 \,\mbox{GeV})\right]
= (3.34 \pm 0.38) \times 10^{-4}\;
\ee
while the next to leading log theoretical number for the SM \cite{sm}
\be\label{br-th}
 Br\left[B \rightarrow X_s + \gamma \;(E_{\gamma} > 1.6 \,\mbox{GeV}) \right]_{\ssc \mathrm{SM}}
= (3.57 \pm 0.30) \times 10^{-4}\;,
\ee
agreeing within 1$\sigma$. Obviously, the only relevant theoretical predictions for a
FCHD $\Phi \to bs$ from any model have to be restricted to parameter space regions
examined to survive the constraint from the radiative $B$ decay.

For the MSSM, the partonic level $b \to s\,\gamma$ transition has
the dominant contributions to the Wilson coefficients of the magnetic and
chromomagnetic penguin operators.  They are induced by the charged current loops
with the W-boson, charged Higgs, and charginos, as well as the gluino and neutralino
loops. CKM-induced flavor violation contributes only to the charged current
loops, which are hence the only directly relevant part
here. Note that the charged Higgs contribution is of the same sign as that of
W-boson, whereas the sign of chargino contributions depend on the sign of the
product $\mu\,A_t$. In the interesting limit of large $\tan\!\beta$, sign of the large
chargino contributions completely dictate the {resulting} branching fraction.
The chargino contributions escape the GIM cancellation through the higgsino components,
with $\tan\!\beta$ enhanced Yukawa couplings. It becomes highly preferable to
maintain a relative negative sign between the $\mu$ and the $A_t$ parameters in
order to bring about cancellations between positive contributions from W-boson
and charged Higgs and negative contribution of chargino loops. That allows the
branching fraction of the radiative decay to fall within the experimentally allowed
range while keeping the Higgs spectrum light.

The FCHD $\Phi \to bs$ for MSSM (with conserved R parity) has been
studied in Refs.\cite{1CT,2CH,3B,5H} (see also
Refs\cite{4H,bejar-LHC}). Refs.\cite{1CT,3B} consider only the
gluino loop contribution. Ref.\cite{1CT} claims branching ratio of
the order of few percents, but it has not put in the \bsg
constraint. The latter implementation has been performed in
Ref.\cite{3B}, with realistic branching fraction in the range of
$10^{-4} - 10^{-3}$ obtained. Much bigger numbers are shown to be
possible but considered unnatural as cancellation among the SM
contribution and the extra, gluino, contribution to  \bsg is
required. In our opinion, the part of the results is of limited
interest not so much because it is  ``unnatural" as it is
unrealistic. After all, the interesting scenarios of the flavor
violation typically involve cancellations among and special
suppression of some ingredients. For instance, we do want
cancellations among the various MSSM contributions to   \bsg, and
large $\Phi \to bs$ is much related to where the dominating
$b\bar{b}$ channel is suppressed. The latter may be considered a
cancellation effect among the various contributions to the effective
coupling. As Ref.\cite{3B} {does not include SUSY electroweak}
contributions, the results relying on a cancellation of the {gluino
part against the others, hence,} have no realistic value for the
full MSSM. {As mentioned above, with flavor off-diagonal soft masses
small enough to remove the dominance of the gluino contribution, the
chargino one with the non-negotiable CKM-induced part becomes very
significant.} The $10^{-4} - 10^{-3}$ numbers {for the branching
fraction results}, however, are already quite encouraging. {More
detailed careful studies are called for.}

In contrast to Refs.\cite{1CT,3B}, Ref. \cite{2CH} concentrates on
the contributions from the electroweak part of the model, obtaining
branching fraction in the range of $10^{-5} - 10^{-3}$. Their
implementation of the \bsg constraint with the help of
micrOMEGAs\cite{micromega}, however, leaves much room for
improvement as the code is based on a minimal flavor violation
scenario as described in Ref.\cite{DnH}.  A more consistent
implementation has been performed in Ref.\cite{5H}.  All the
{published} studies other than Ref.\cite{5H} take as the source of
flavor violation the $LL$-mixings in squark masses of the second and
third generations. While {for most of the models on SUSY
breaking/mediation} the $LL$-mixings is commonly expected to be
larger than $LR$- or $RR$-mixings, {purely phenomenological results
on} the latter, {independent of the SUSY breaking models,} are of
interest in their own right. Moreover, the argument that
$LR$-mixings are of less interest because they are more stringently
constrained by  \bsg  does not stand. The $LR$-mixings are more
constrained because they contribute more effectively to  \bsg ,
which requires a chiral flip. Likewise, they contribute more
effectively to $\Phi \to bs$. As mentioned above, the $LR$-mixings
means flavor off-diagonal $A$-terms and hence Higgs-sfermion
vertices. As it is the admissible $\Phi \to bs$ surviving the  \bsg
constraint that we are after, the $LR$-mixings, or rather flavor
off-diagonal $A$-terms, do not clearly appear to be at a position of
disadvantage when compared to the $LL$-mixings soft mass terms. This
is especially relevant with the gluino contribution.

It is interesting to see different parts of the contributions to
$\Phi \to bs$ within MSSM. However, {having the electroweak
contribution without the gluino one} is not realistic. The more
interesting question there is if the electroweak contribution could
dominate over the gluino one. The limiting case of that is the
minimal CKM-induced flavor violation scenario on which we will
report {explicit} results below. More detailed comprehensive studies
for the generic {case is certainly} in order. Ref.\cite{5H} is an
attempt in the direction. However, the treatment of the up- and
down-sector $LL$-mixings as independent or identical parameters is
not exactly correct. All discussions about the electroweak
contributions have not separated the CKM-induced part from that
which has the soft SUSY parameters as its source. {We take up the
task here, as a step to reveal more interesting details of the
FCHDs.} Our result here that even the purely CKM-induced
contribution can get to the $10^{-4} - 10^{-3}$ level would likely
be beyond expectations.

For all the numerical results reported below, we switch off all soft sector flavor violations.
The structure of the MSSM results is well known and the analytical formulas involved in
the calculation are by now standard. We have no disagreement with previous authors on
the matter. Hence, unless where necessary for the illustration of the issues under
discussion, we will mostly refrain from showing such expressions in this brief letter
but only highlight the more important points involved and give relevant references.
Note that we stick to full mass eigenstate formalism, no mass-insertion approximation applied.

We have used for the $B \to X_s + \gamma$ calculation the NLO expressions given in
Ref.\cite{NK}, while for the Wilson coefficients $C_7$ and $C_8$ we have used the
expression given in Ref.\cite{c7c8} at LO in the framework of MSSM with CKM
as the only source of flavor violation. We impose the constraint asking the theoretical
number to lie within the 3$\sigma$ experimental bounds.

We have computed Higgs decay width for $\Gamma_{\ssc\!\Phi}$ for all the three
mass eigenstates $h^{\ssc 0}$, $H^{\ssc 0}$, or $A^{\ssc 0}$, including the
following partial widths at the leading order :
\begin{eqnarray}
\Gamma_{\ssc\!\Phi} &=&\sum_{f}\Gamma(\Phi\to f\bar{f}) +
\Gamma(\Phi\to VV)
\nonumber \\
&& +  \Gamma(\Phi\to V H_i)
+\Gamma(\Phi\to H_i H_j)+\Gamma(\Phi\to {\rm SUSY \, particles})\label{widd}
\end{eqnarray}
with expressions in accordance with Ref.\cite{D}. QCD corrections
to $\Phi \to f\bar{f}$ and  $\Phi \to \{V^*V^*, VV^*, V^*H_i\}$
decays are not included in the width. Since the width of $\Phi\to {b}s$ may
become comparable to $\Phi \to \{gg, \gamma \gamma\}$,
it is necessary to include $\Phi \to \{ g g, \gamma \gamma\}$
in the computation as we have done.
{All the computations of the rate for FCHD are
done with the help of the packages {FeynArts,FormCalc}~\cite{FA2}, and with LoopTools
and FF for numerical evaluations~\cite{FF,LT}. The one-loop amplitudes are evaluated in
the 't Hooft--Feynman gauge using dimensional regularization.}

The MSSM Higgs sector is parametrized by the mass of the CP-odd $M_{\!\ssc A}$ and
$\tan\!\beta$ while the top quark mass and the associated squark masses enter
through radiative corrections \cite{okada}. In our study we include the leading
corrections only, {with the Higgs masses} given by
\begin{eqnarray}
& & m^2_{\!\ssc h^0\!\!,H^0} =  \frac{1}{2}\Big[ M_{\!\ssc AZ}^2
\mp \sqrt{ M_{\!\ssc  A\!Z}^4-4 M_{\!\ssc A}^2 M_{\!\ssc Z}^2\, {\cos}^2 2\beta
-4\epsilon M_{\!\ssc A}^2 \,\sin\!^2\!\beta +M_{\!\ssc Z}^2 \, \cos\!^2\!\beta } \Big] \\
& &\tan \!2 \alpha
=  \frac{M_{\!\ssc A}^2 + M_{\!\ssc Z}^2}{M_{\!\ssc A}^2 - M_{\!\ssc Z}^2 +
\epsilon/ \cos\! 2\beta }
\, \tan\! 2 \beta\;,
\quad\qquad  \,-\frac{\pi}{2} \le \alpha \le 0\, \; ,
\end{eqnarray}
where
\begin{eqnarray}
& & M_{\!\ssc A\!Z}^2 = M_{\!\ssc A}^2+M_{\!\ssc Z}^2+\epsilon \; ,
\qquad\qquad
\epsilon = \frac{3 G_{\!\ssc F}}{\sqrt{2} \pi^2 } \frac{m_t^4}{ \sin\!^2\!\beta}
\mbox{log} \left[\frac{m_{\tilde{t}_1}m_{\tilde{t}_2}}{m_t^2} \right] \;.
\end{eqnarray}

\begin{figure}[t]
\centerline{
\epsfxsize2.99 in
\epsffile{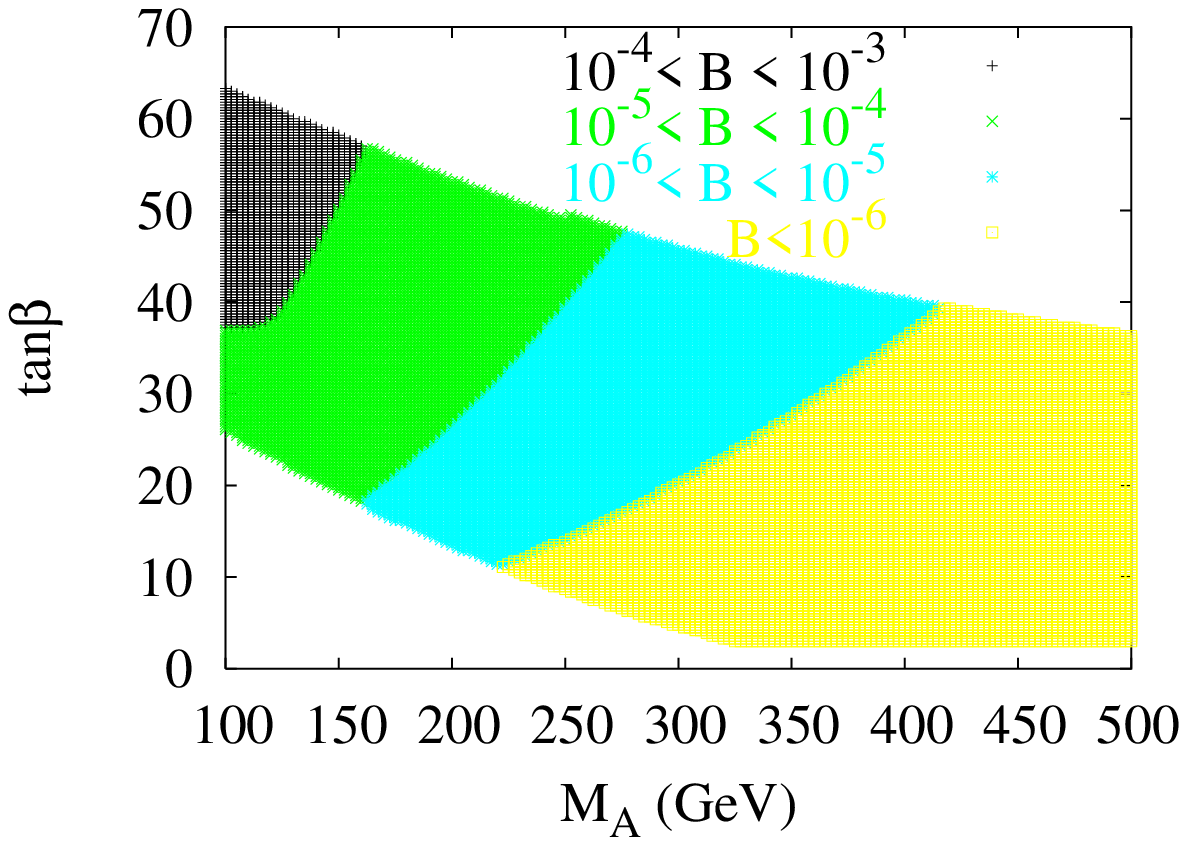}
\hskip0.4cm
\epsfxsize2.99 in
\epsffile{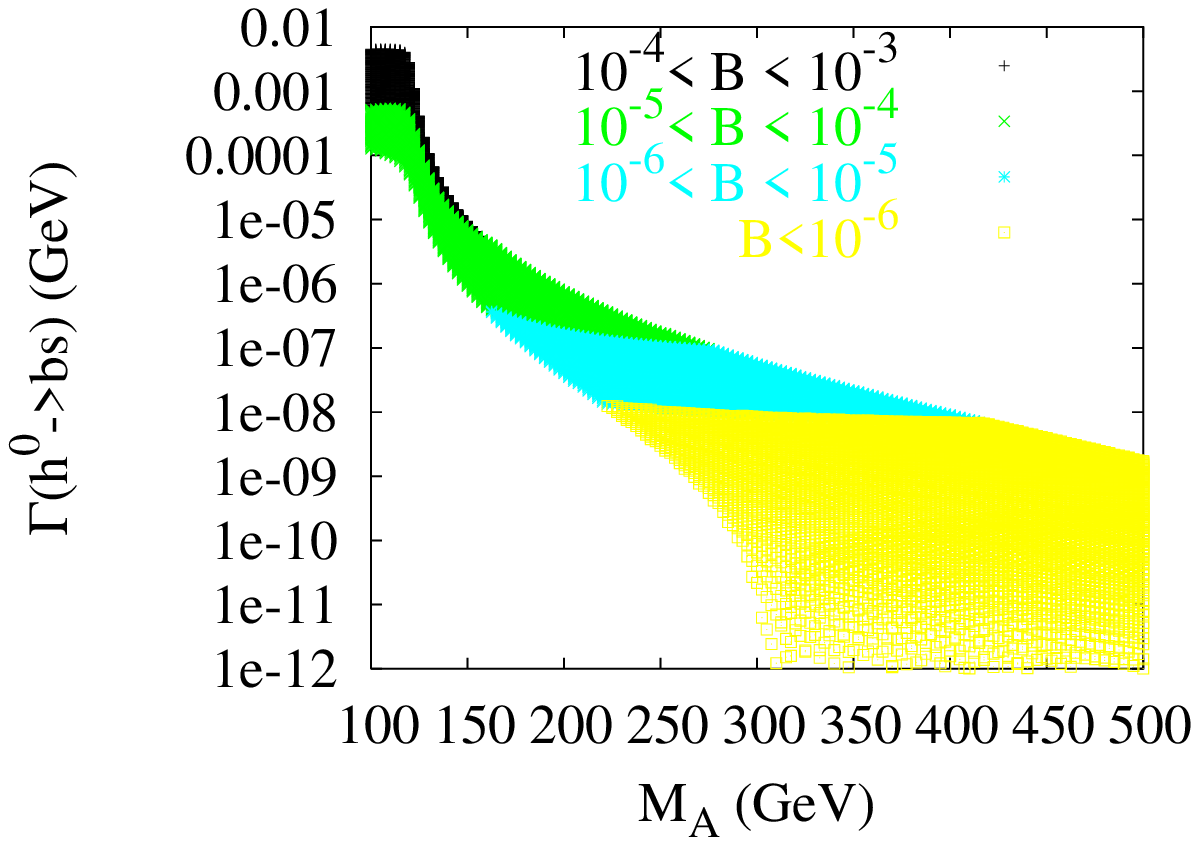}}
\smallskip\smallskip
\caption{Branching ratio of $h^{\ssc 0}\to bs$ in the plane of
$M_{\!\ssc A},\tan\!\beta$ (left panel) and the decay width $\Gamma(h^{\ssc 0}\to bs)$
in GeV (right panel) as a function of $M_{\!\ssc A}$ with $2<\tan\!\beta<65$,
  for  $M_{\mbox{\tiny SUSY}}=800\,\mbox{GeV}$, $\mu=1 \,\mbox{TeV}$,
$ A_t=A_b=A_{\tau}=-1.6\,\mbox{TeV}$, and $M_2=200\,\mbox{GeV}$.
From up-left to down-right $B=B(h^{\ssc 0}\to bs)$ is: $10^{-4}<B<10^{-3}$,
$10^{-5}<B<10^{-4}$, $10^{-6}<B<10^{-5}$, $B<10^{-6}$.
}
\label{fig1}
\end{figure}

\begin{figure}[t]
\centerline{
\epsfxsize2.99 in
\epsffile{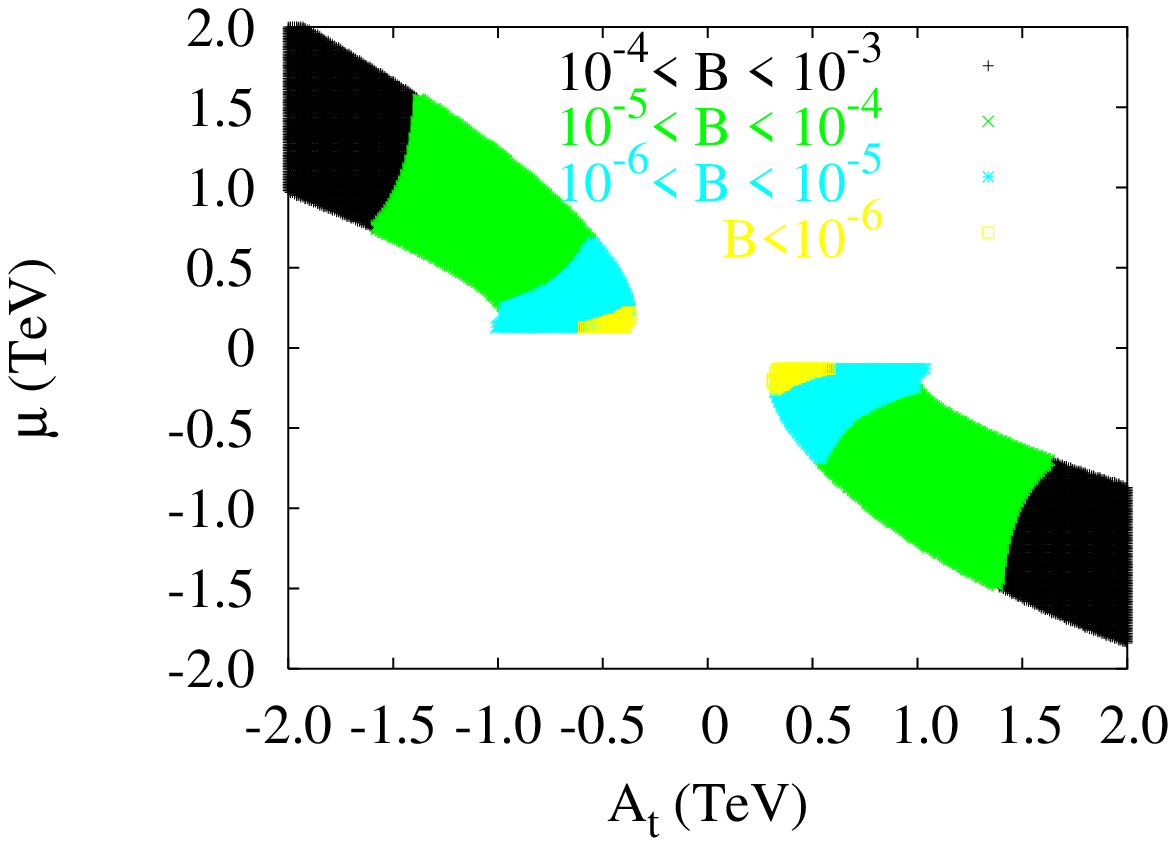}
\hskip0.4cm
\epsfxsize2.99 in
\epsffile{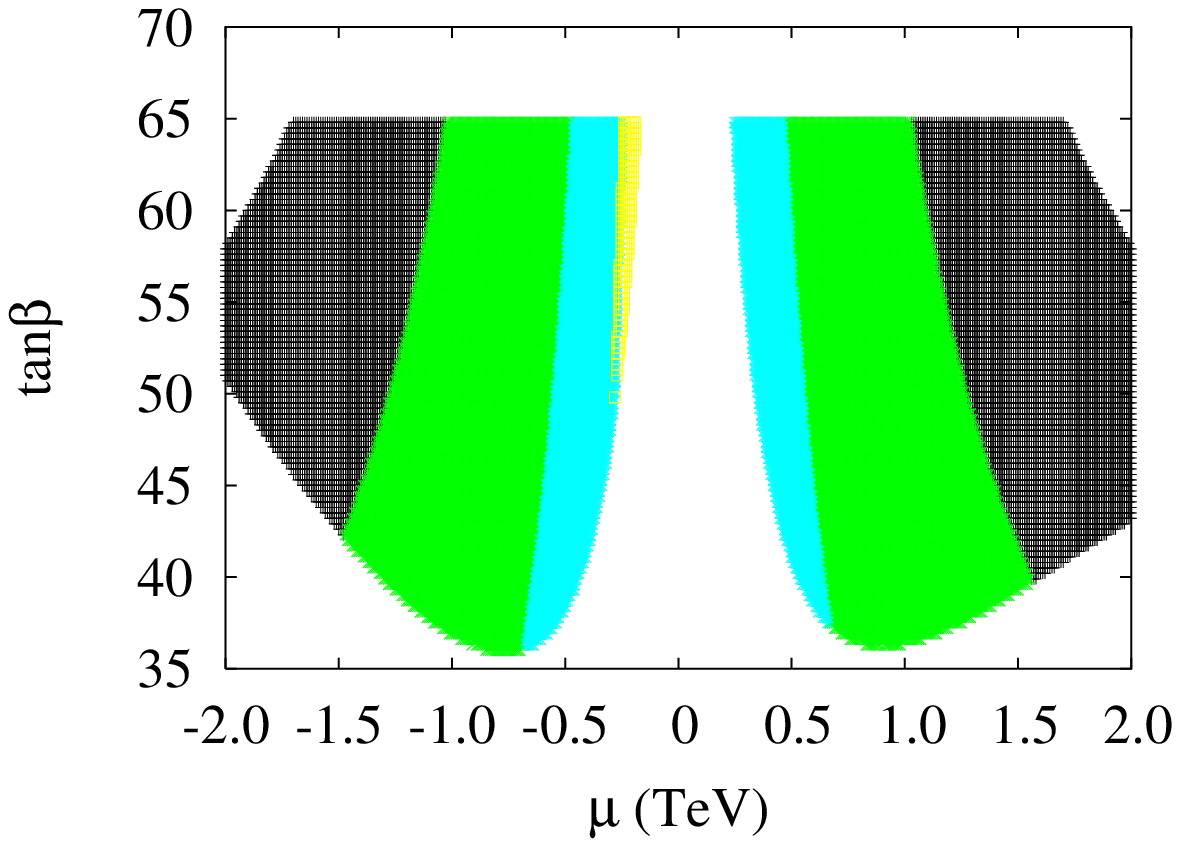}}
\smallskip\smallskip
\caption{Branching ratio of $h^{\ssc 0}\to bs$ in the $(A_t,\mu)$ plane
with $\tan\!\beta=45$ (left panel) and $(\mu,\tan\!\beta)$ with $A_t=-\mu$ (right panel)
 with $M_{\!\ssc A}=130\,\mbox{GeV}$, $M_{\mbox{\tiny SUSY}}=800\,\mbox{GeV}$,
and $M_2=200\,\mbox{GeV}$. From left to right or right to left: $10^{-4}<B<10^{-3}$,
$10^{-5}<B<10^{-4}$, $10^{-6}<B<10^{-5}$, $B<10^{-6}$.}
\label{fig2}
\end{figure}

Our major result is illustrated in Fig.\ref{fig1}, in which we plot
the branching fraction of $h^{\ssc 0} \to bs$ in the plane
($M_{\!\ssc A},\tan\beta$) (left panel) and the decay width in GeV
as a function of $M_{\!\ssc A}$ (right panel) for $M_{\mbox{\tiny
SUSY}}=800\,\mbox{GeV}$, {$\mu=1 \,\mbox{TeV}$}, {and a common value
for the $A$-terms ($A_t=A_b=A_{\tau}$) at} $-1.6\,\mbox{TeV}$
($=-{\rm sign}(\mu)\, 2 M_{\mbox{\tiny SUSY}}$), and
$M_2=200\,\mbox{GeV}$; $M_{\mbox{\tiny SUSY}}$ is the common soft
mass for all squarks involved. Note that we take $A_t$ with a
relative negative sign to the $\mu$-parameter as preferred for \bsg.
As also noted in Ref.\cite{3B}, large branching fraction for
$h^{\ssc 0} \to bs$ is obtained in the parameter range of the small
$\alpha_{\mbox{\tiny eff}}$ scenario\cite{sea}. As $\tan\!\beta$
runs here, it tunes the loop corrections to the Higgs mass, {\em
i.e.} it tunes  $\alpha_{\mbox{\tiny eff}}$. {Small  $M_{\ssc\! A}$
and large $\tan\!\beta$, and positive $\mu$ are preferred.} As one
can see from left plot, large branching fraction ${\cal  B}>10^{-5}$
is obtained for large $\tan\!\beta\ga 30$ and $M_{\ssc\!
A}<250\,\mbox{GeV}$. The {unmarked regions in the plots are excluded
by the violations of} either $b\to s\gamma$, $\delta\rho$, or one of
the experimental limit on neutralino and chargino masses.

In Fig.\ref{fig2}, we show the branching fraction of $h^{\ssc 0}\to
bs$ in the plane of $(A_t,\mu)$ with $\tan\!\beta=45$ (left panel)
and $(\mu,\tan\beta)$ with $A_t=-\mu$ (right panel), {respectively,}
for $M_{\ssc\! A}=130\,\mbox{GeV}$, $M_{\mbox{\tiny SUSY}}=800$ GeV,
$M_2=200\,\mbox{GeV}$. It is clear that large branching fraction is
obtained for large $A_t$, large $\mu$ as well as large
$\tan\!\beta$. Similar to Fig.\ref{fig1},  {unmarked regions in the
plots are excluded by the violations of} either $b\to s\gamma$,
$\delta\rho$, or one of the experimental limit on neutralino and
chargino masses. We mention in passing that from the left plot one
can see that $b\to s\gamma$ favors $\mu A_t$ to be negative.

\begin{figure}[t]
\centerline{
\epsfxsize2.99 in
\epsffile{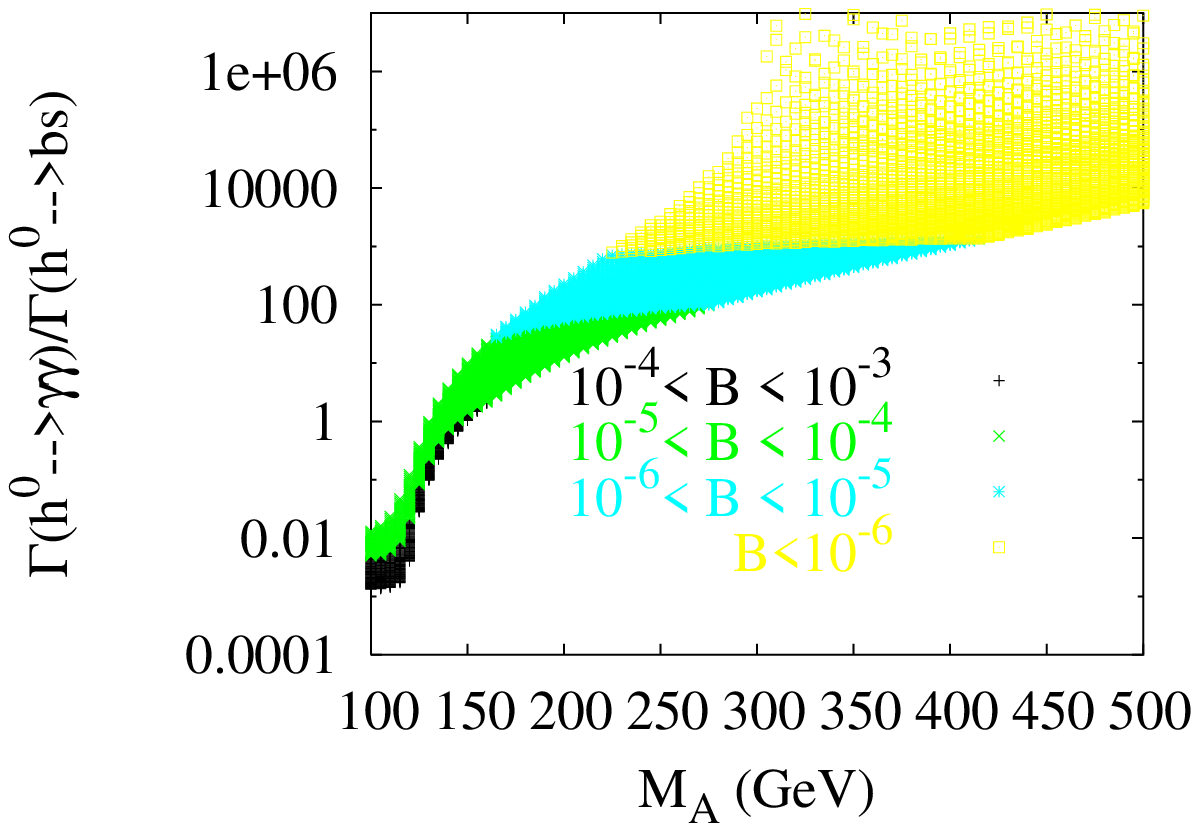}
\hskip0.4cm
\epsfxsize2.99 in
\epsffile{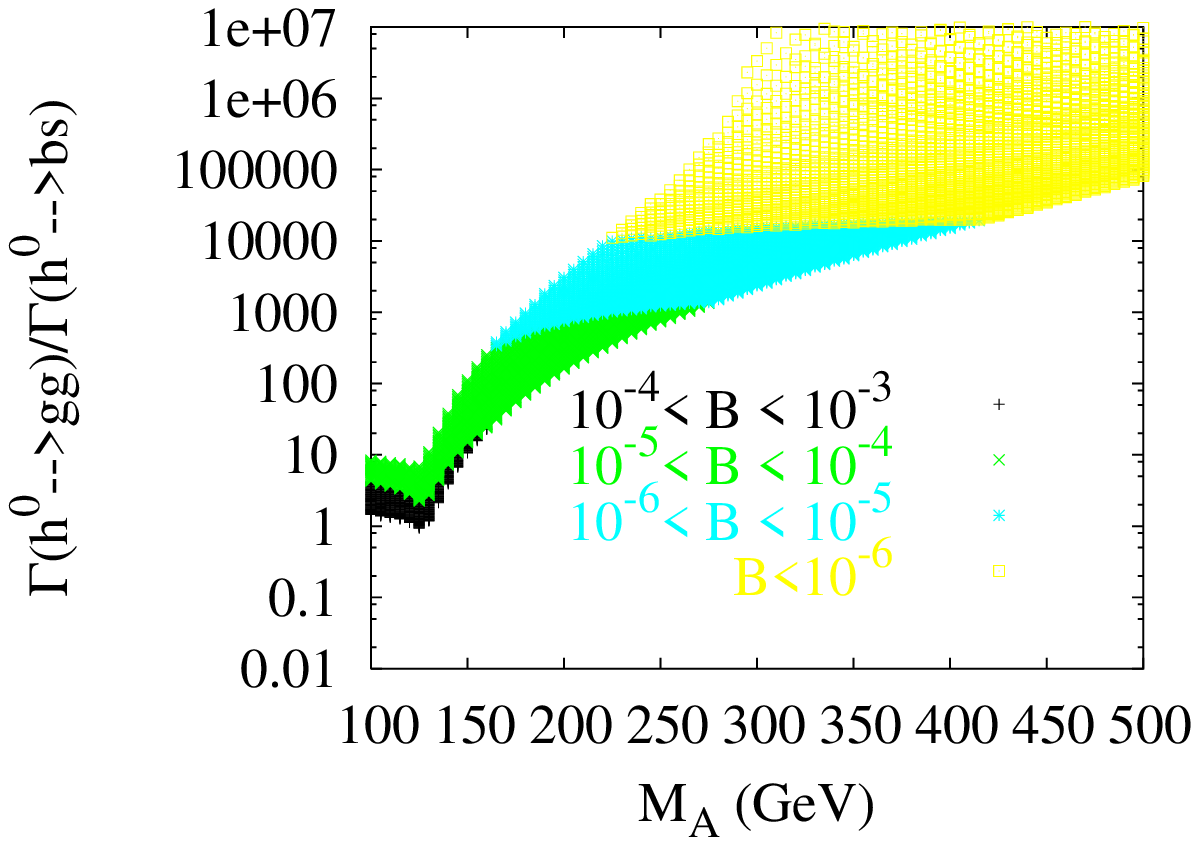}}
\smallskip\smallskip
\caption{Ratio of $\Gamma(h^{\ssc 0}\to \gamma\gamma)/\Gamma(h^{\ssc 0}\to bs)$
(left panel) and $\Gamma(h^{\ssc 0}\to gg)/\Gamma(h^{\ssc 0}\to bs)$ (right panel)
as a function of $M_{\!\ssc A}$ with $2<\tan\!\beta<65$. The other parameters
same as for Fig.\ref{fig1}.
From down-left to up-right $B=B(h^{\ssc 0}\to bs)$ is: $10^{-4}<B<10^{-3}$,
$10^{-5}<B<10^{-4}$, $10^{-6}<B<10^{-5}$, $B<10^{-6}$.
}
\label{fig3}
\end{figure}


\begin{figure}[t]
\centerline{
\epsfxsize2.99 in
\epsffile{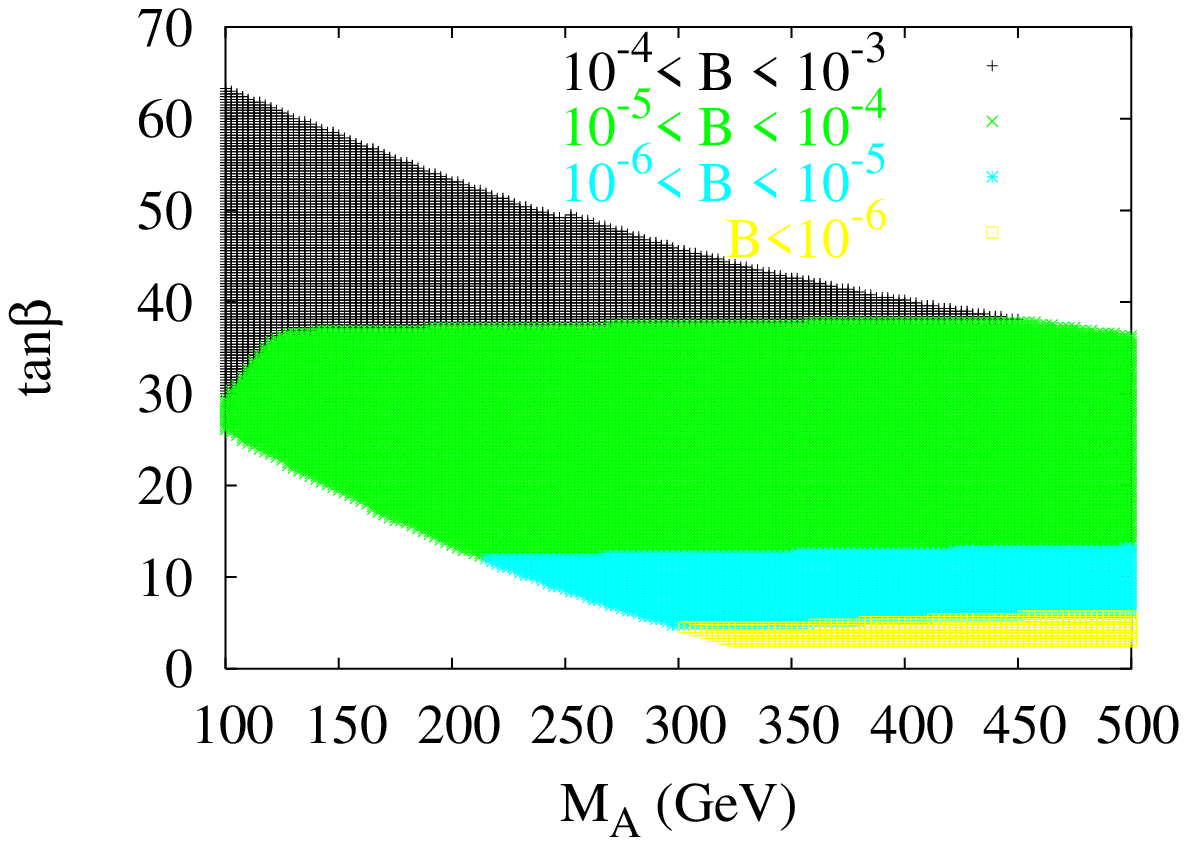}
\hskip0.4cm
\epsfxsize2.99 in
\epsffile{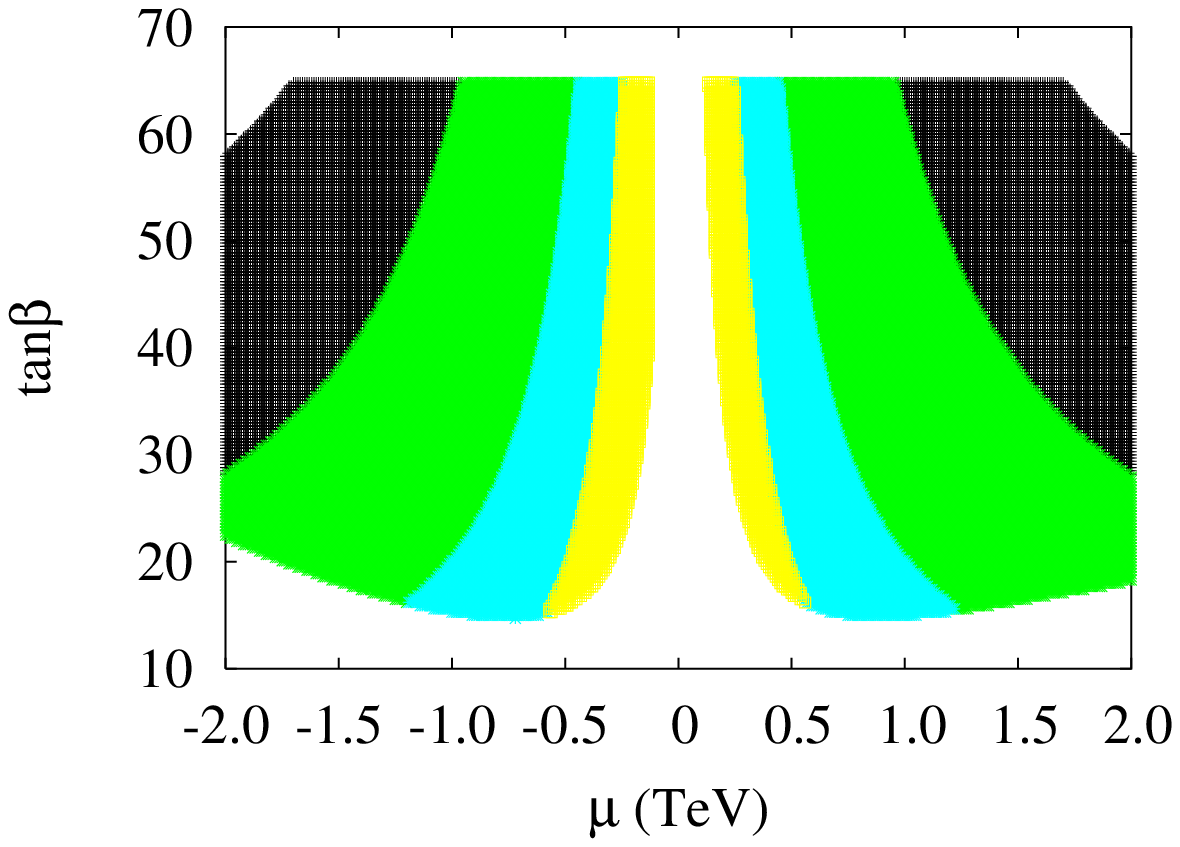}}
\smallskip\smallskip
\caption{Branching ratio of $H^{\ssc 0}\to bs$ in the
$M_{\!\ssc A},\tan\!\beta$ plane (left panel) and $(\mu,\tan\!\beta)$ plane (right panel),
the former with $\mu=1\,\mbox{Tev}$, the latter  with $M_{\!\ssc A}=250\,\mbox{GeV}$,
$A_t=-\mu$. The other unspecified parameters are same as for Fig.\ref{fig1}.
Left panel: from up-left to down-right $B=B(H^{\ssc 0}\to bs)$ is: $10^{-4}<B<10^{-3}$,
$10^{-5}<B<10^{-4}$, $10^{-6}<B<10^{-5}$, $B<10^{-6}$.
Right panel: From left to right or right to left: $10^{-4}<B<10^{-3}$,
$10^{-5}<B<10^{-4}$, $10^{-6}<B<10^{-5}$, $B<10^{-6}$.
}
\label{fig4}
\end{figure}


We illustrate in Fig.\ref{fig3}, the ratios {
$\Gamma(h^{\ssc 0}\to \gamma\gamma)/(\Gamma(h^{\ssc 0}\to bs))$ (left panel)
and $\Gamma(h^{\ssc 0}\to gg)/(\Gamma(h^{\ssc 0}\to sb))$ (right panel)} as a
function of $M_{\ssc\! A}$ for  {$2<\tan\beta<65$, and  otherwise the
same model parameters} as in Fig.\ref{fig1}. From the left plot one can read that
$\Gamma(h^{\ssc 0}\to bs)$ can be both larger or smaller than
$\Gamma(h^{\ssc 0}\to \gamma \gamma)$ while from the right plot we can see that
$\Gamma(h^{\ssc 0}\to gg)$ is almost all the time larger than $\Gamma(h^{\ssc 0}\to bs)$.
The $h^{\ssc 0} \to bs$ partial width is in fact comparable to that of
$h^{\ssc 0} \to \gamma\gamma$, as we illustrated in Fig.\ref{fig3}.

Situation for the heavier CP even state $H^{\ssc 0}$ is similar, as
illustrated in Fig.\ref{fig4}, based on the same set of parameter
input. As in the case of  {$h^{\ssc 0}$}, the branching fraction of
$H^{\ssc 0}\to bs$ can reach $10^{-4}$ level. The corresponding
result for the pseudoscalar is very similar to  {$H^{\ssc 0}$} and
we do not show it here.

At this point, some comments on the issues related
to the limitations of the approximation used in our calculations are
in order. It has been well appreciated that there are important QCD
corrections to the Higgs to fermion decay widths as well as significant
(resummed) large $\tan\!\beta$ corrections. We have not implemented the
corrections into our calculations here, for the main reason that a
consistent implementation is very involved, in fact quite beyond
the scope of present work or what has been done by other 
authors on this topic so far.
Moreover, we believe the present analysis does illustrate at least
qualitatively some interesting physics which is not going to be
totally invalidated by a complete consistent higher order calculations
incorporating the corrections. While how to implement the above
corrections to the, flavor-conserving Higgs decays is well
known and even built into the standard numerical codes, a consistent
implementation here requires the simultaneous implementation onto the
flavor changing $\Phi \to bs$ part as well as the \bsg background
constraint. A partial implementation, say onto flavor-conserving Higgs
decays only, will likely give misleading results --- a point also
made in Ref.\cite{3B}. For example, QCD correction is known to reduce
the Higgs to $b\bar{b}$ width.  However, it will more or less have a
similar effect on the $\Phi \to bs$ width. In fact, one may expect
some partial cancellation between the two rendering the overall effect
on the branching ratio we are interested in less than what the
correction to Higgs to $b\bar{b}$ in itself may indicate.  
And there is still the corresponding corrections to \bsg to worry
about.  The situation for the large $\tan\!\beta$ corrections is
similar. Of course part of the corrections involved may be somewhat
sensitive to the values of the SUSY parameters that have otherwise
minimal role to play here --- the gluino mass is one clear example. 
Thus, to the leading order, our results are consistent and not inferior to
the results reported in the previous works. Rather the effect of CKM
induced contributions have been brought out more clearly.

  We expect our conclusion to qualitatively remain valid over quite
  some region of the general SUSY parameter space. Let us take a look
  at the plausible numerical impact of the corrections here, focusing
  on the small Higgs mass region where they have a strong effect. The
  region is also where our results look most interesting. QCD
  correction reduces the Higgs to $b\bar{b}$ (and other fermion
  channels) width. Our focus here being the branching ratio of the
  flavor changing decays, within the constraint of flavor violation
  admissible in \bsg . Had we naively included the QCD corrections to
  the Higgs to $b\bar{b}$, we may have ended up in false enhancement.
  The (resummed) large $\tan\!\beta$ correction is more relevant here,
  as our results give interesting branching ratio mostly in the large
  $\tan\!\beta$ region. The $\tan\!\beta$ effect is, in first order, a
  correction on the $b$-Yukawa coupling.  Depending on the SUSY
  parameters, the effect can reach a $60$-$70\%$ enhancement of the
  (like $b\bar{b}$) width. However, a substantially reduced effect due
  to partial cancellation of various contributions is not
  inconceivable. As the major contribution to the $\Phi \to bs$
  channel discussed here comes from the chargino-stop diagram, the
  large $\tan\!\beta$ correction to the flavor changing width would be
  of higher order. However, there is still the effect on the \bsg
  constraint to be considered. On the whole, we think it is reasonable
  to expect that the correction would not change the admissible branching
  ratio by an order of magnitude.


\begin{figure}[t]
\centerline{
\epsfxsize2.99 in
\epsffile{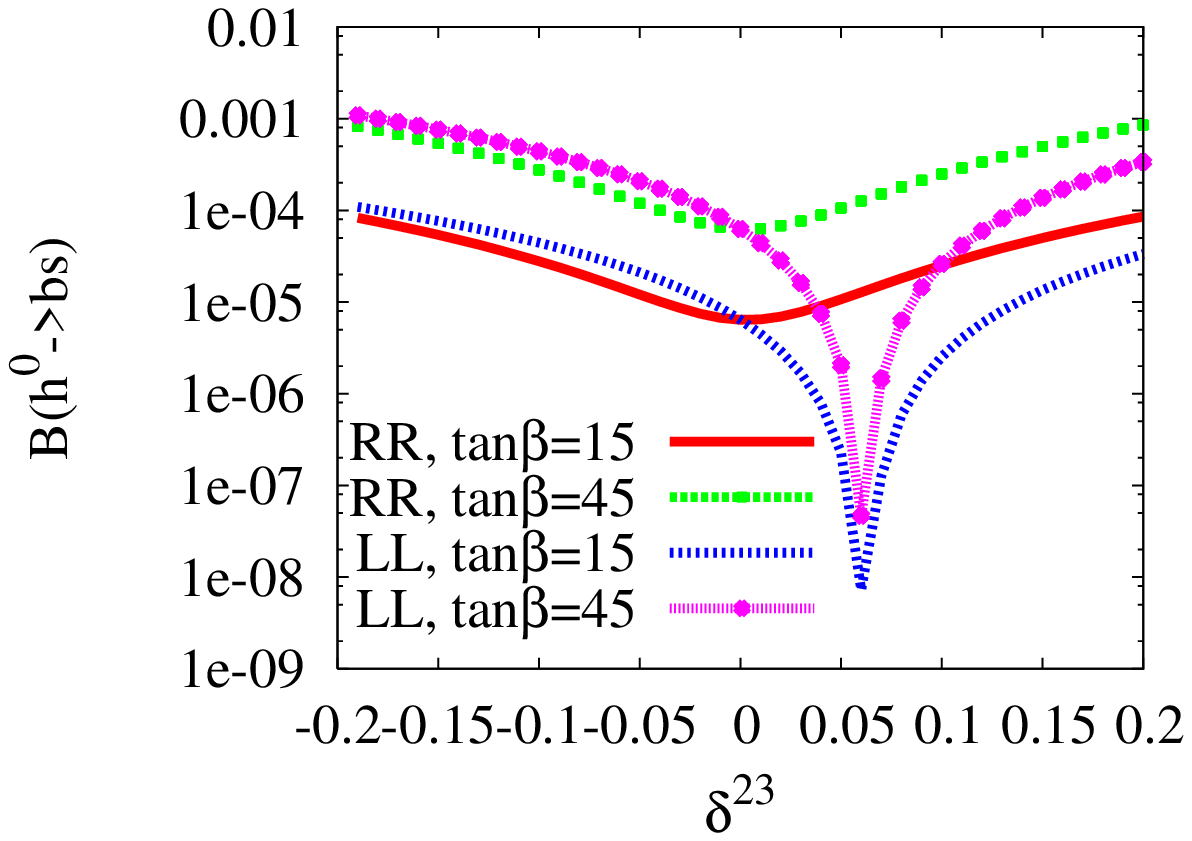}
\hskip0.4cm
\epsfxsize2.99 in
\epsffile{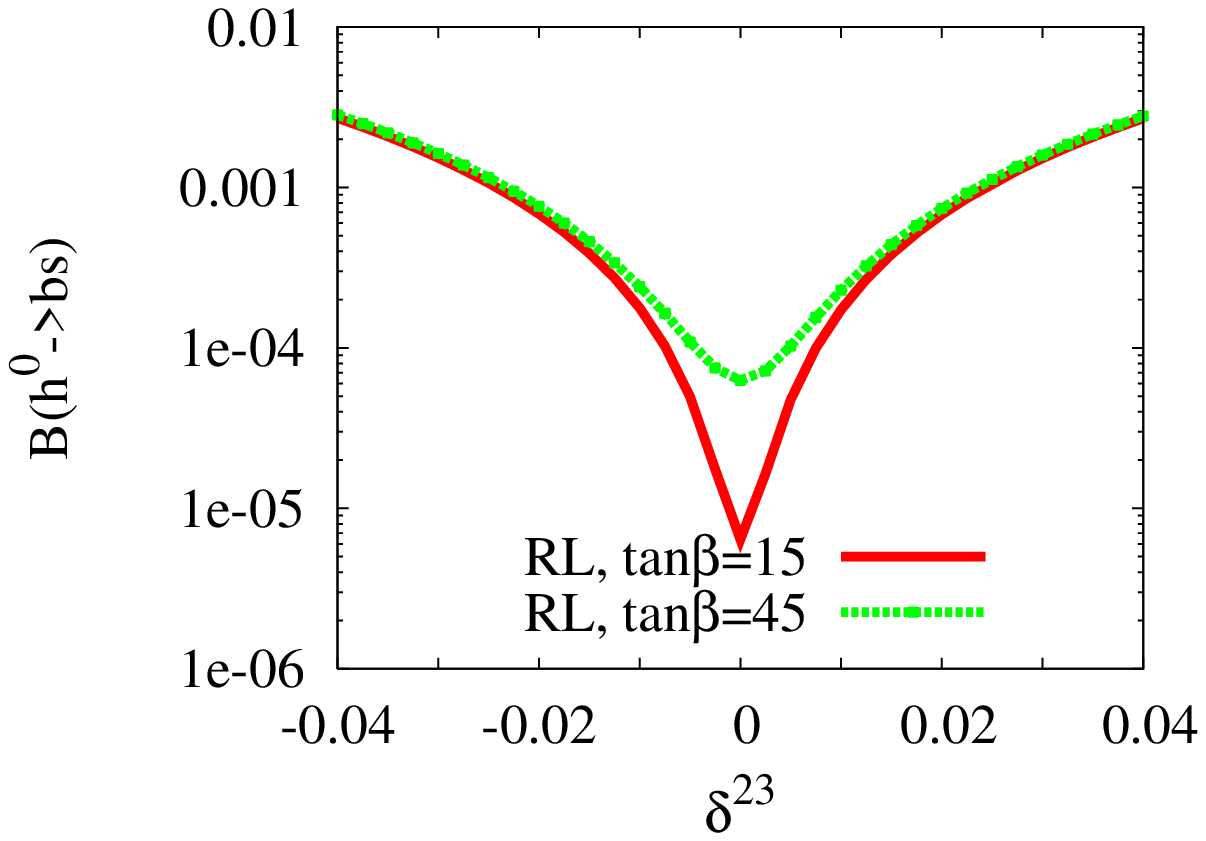}}
\smallskip\smallskip
\caption{Naive $(h^{\ssc 0}\to bs)$ branching ratio results
with flavor violating soft mass inputs, for comparison.
The other parameters same as for Fig.\ref{fig1}.
}
\label{fig6}
\end{figure}

We further illustrate some naive results on the $h^{\ssc 0}\to bs$
branching ratios with flavor violating soft mass inputs to give the
readers a direct comparison. In Fig.\ref{fig6}, the branching ratios
for the, separate, cases with nonzero $LL$-, $RR$-, or $RL$- (flavor off-diagonal
$A$-term) mixings among the second and third family squarks are illustrated,
for otherwise the same model parameters. Two specific values of $\tan\!\beta$,
15 and 45, are chosen for clear illustration. Notice that the minima of the curves
apart from the $LL$ cases are at zero mixings, with value given by the
independent CKM induced contribution. For the cases of the $LL$-mixings,
The CKM induced effect cannot be separated as an independent part. In fact,
the minima are off-set from zero mixings and go deeper as the CKM induced
effect is partially canceled. The pairs of curve for the two $\tan\!\beta$
values also illustrate the basic trend of dependence on the parameter. The
$LL$ and $RR$ cases have the $\tan\!\beta$ enhancement as for the CKM only
results. For the $RL$ case, however, as it is actually a linear combination
of the soft ($A$-) term and the $\tan\!\beta$-dependent $F$-term that matters,
$\tan\!\beta$-dependence fades out at large ($A$-term) mixings. The plots are
shown only for the purpose of illustrating the basic theoretical feature.
Hence, no checking against \bsg or other constraints applied.


In summary, we have discussed the potentially experimentally
interesting flavor changing Higgs decays $\Phi \to bs$ for the MSSM,
and presented explicit numerical results focusing on the minimal
flavor violation scenario. The scenario has FCNC induced through
mostly chargino loops with the CKM mixing as the sole, and
unavoidable, source of flavor violation.  We illustrate that the
branching fractions for the decays of all three Higgs states, after
factoring in the other major experimental constraints such as \bsg,
can still reach the range of $10^{-4} - 10^{-3}$, without any extra
flavor violation of SUSY origin. The numbers are comparable with
that obtained by the other authors through assuming flavor violating
soft squark masses. Our calculations have not
incorporated the major QCD and large $\tan\!\beta$ corrections, the
consistent implementation of which still has to be performed. We
explained the related concerns and our belief that the interesting
branching ratio illustrated is very unlikely to be altered by  an
order of magnitude.

{\it Acknowledgements:} We would like to thank F. Borzumati, 
Y. Hsiung and G. Mazumdar for
helpful discussions.  O.K. is partially supported by research
grants number 94-2112-M-008-009 and 95-2112-M-008-001 from the NSC
of Taiwan.


\begin{thebibliography}{99}
\bibitem{bob}
{The problem here is with $s$-jet identification, as versus $b$- as well as
$u$- and $d$-jets, which is very difficult but not impossible. --- Y. Hsiung, private communiation;
G. Mazumdar, private communiation.}
\end{thebibliography}
\end{document}